\documentclass[%
 aip,
 apl,%
 amsmath,amssymb,
preprint,
]{revtex4-1}

\usepackage{graphicx}
\usepackage{dcolumn}
\usepackage{bm}
\usepackage{subfig}
\usepackage{floatrow}

\def\k{\mathbf{k}}
\def\p{\mathbf{p}}

\begin{document}


\title[the title]{Large g factor in bilayer WS$_2$ flakes}

\author{Sibai Sun}
\author{Yang Yu}
\author{Jianchen Dang}
\author{Kai Peng}
\author{Xin Xie}
\author{Feilong Song}
\author{Chenjiang Qian}
\author{Shiyao Wu}
\author{Hassan Ali}
\affiliation{Beijing National Laboratory for Condensed Matter Physics, Institute of Physics, Chinese Academy of Sciences, Beijing 100190, China}
\affiliation{CAS Center for Excellence in Topological Quantum Computation and School of Physical Sciences, University of Chinese Academy of Sciences, Beijing 100049, China}%
\author{Jing Tang}
\affiliation{Laboratory of Quantum Engineering and Quantum Metrology, School of Physics and Astronomy, Sun Yat-Sen University (Zhuhai Campus), Zhuhai 519082, China}
\author{Jingnan Yang}
\author{Shan Xiao}
\author{Shilu Tian}
\author{Meng Wang}
\affiliation{Beijing National Laboratory for Condensed Matter Physics, Institute of Physics, Chinese Academy of Sciences, Beijing 100190, China}
\affiliation{CAS Center for Excellence in Topological Quantum Computation and School of Physical Sciences, University of Chinese Academy of Sciences, Beijing 100049, China}%
\author{Xinyan Shan}
\affiliation{Beijing National Laboratory for Condensed Matter Physics, Institute of Physics, Chinese Academy of Sciences, Beijing 100190, China}
\author{M. A. Rafiq}
\affiliation{Department of Physics and Applied Mathematics, Pakistan Institute of Engineering and Applied Sciences, P.O. Nilore Islambad 45650, Pakistan}
\author{Can Wang}
\author{Xiulai Xu}%
\email{xlxu@iphy.ac.cn}
\affiliation{Beijing National Laboratory for Condensed Matter Physics, Institute of Physics, Chinese Academy of Sciences, Beijing 100190, China}
\affiliation{CAS Center for Excellence in Topological Quantum Computation and School of Physical Sciences, University of Chinese Academy of Sciences, Beijing 100049, China}
\affiliation{Songshan Lake Materials Laboratory, Dongguan, Guangdong 523808, China}

\date{\today}

\begin{abstract}
The valley of transition metal dichalcogenides provides an additional platform to manipulate spin due to its unique selection rule. Normally, intralayer optical transitions in magnetic field show a Zeeman splitting with g factor of about $-4$. Here we report remarkable valley Zeeman effect exhibited by splitting of excitonic emission in a bilayer WS$_{2}$, with a value of g factor as large as $-16.5$. The observed large g factor results from the interlayer recombination, as the conduction band and valence band are modified in opposite directions by magnetic field in different layers. The interlayer recombination is due to the defect induced inversion symmetry breaking, which is theoretically not accessible in ideal bilayer WS$_{2}$ with inversion symmetry. Large g factor of interlayer emission offers potential benefits for future optical spin control and detection.

\end{abstract}

\maketitle

\floatsetup[figure]{style=plain}

Monolayer transition metal dichalcogenides (TMDs) have attracted considerable interests in their potentials for the next generation of nano-devices based on the valley pseudospin of electrons and holes\cite{xu_review}. Valley-dependent optical polarization, arising from inversion symmetry breaking and strong spin-orbit coupling, provides a platform to manipulate spin information of photon, thus having prospects in quantum information processing\cite{xu_bilayerQuantumGates,singlePhoton_WSe2_ymh,singlePhoton_WSe2_ajit}. The study of the Zeeman splitting of the excitonic states plays an important role in investigating valleytronics, in which Land\'{e} g factor about $-4$ has been obtained in monolayer WS$_{2}$ both theoretically\cite{mac_gFactorMonolayer} and experimentally\cite{jan_magopt,andreas_magopt,korn_gFactorMonolayer}.

Recently, Giant pseudospsin g factors up to 38 have been observed in monolayer MoSe$_2$ with controlling carrier concentration using a field transistor structure \cite{PhysRevLett.120.066402,gFactor38_MoSe2,Gustafsson2018}. In the bilayer TMDs, however, there is an opportunity to increase the Zeeman splitting greatly by inducing interlayer emission. In natural TMDs materials, the adjacent layers are in reverse directions on the layer plane. The interlayer emission photons from adjacent layers are from combinations of electrons and holes in different valleys. Their corresponding valley orbital angular momenta are thus in opposite directions. When a magnetic field is applied, the energy levels of electrons and holes shift to opposite directions. Thus a larger g factor could be observed. Unfortunately, in ideal pristine bilayer TMDs, the interlayer emission is suppressed due to inversion symmetry\cite{xu_bilayerQuantumGates,cui_forbidInterlayer}. Once inversion symmetry is broken, interlayer emission is no longer suppressed and a larger Zeeman splitting will appear. In fact, large g factors induced by symmetry breaking have been reported in heterostructures in MoSe$_2$ /WSe$_2$ \cite{korn_heteroLargeG} and in a bilayer MoTe$_2$ \cite{gao_gFactor}. However, large g factor has not been reported in a bilayer WS$_2$.

In this work, we report a large g factor in bilayer WS$_2$ with numerous defects. The WS$_2$ flakes were fabricated by exfoliating and the bilayer was confirmed by the thickness and the Raman characteristics. The Zeeman effect was investigated by the magneto-photoluminescence spectroscopy with a large g factor of about 16.5 observed at 30 K. We propose that the defect induced symmetry breaking leads to the interlayer emissions, resulting in the largely increased g factor.

\begin{figure}%
\centering
\includegraphics[width=\columnwidth]{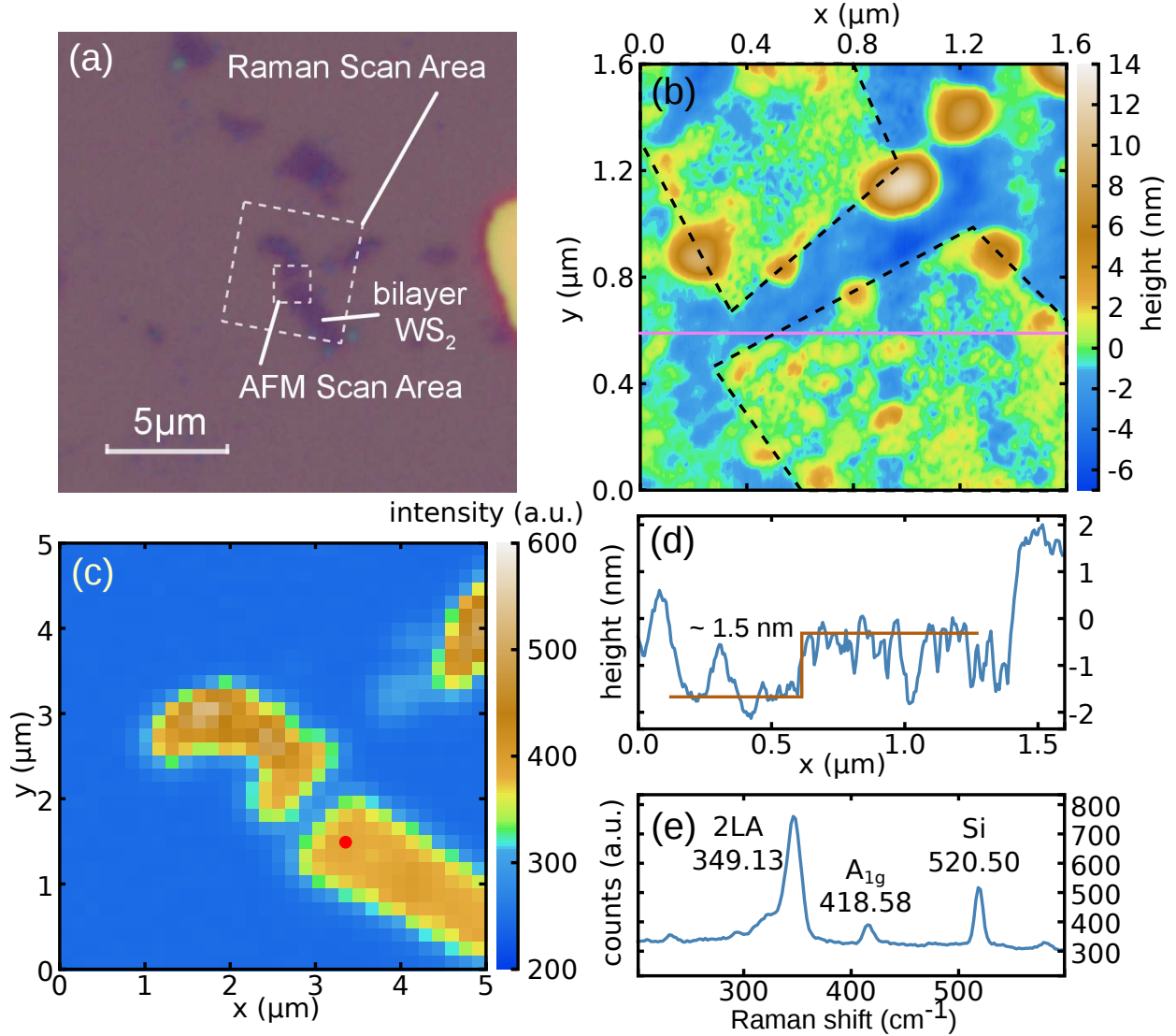}
\caption[]{\label{fig:geo}%
\raggedright
(a) Optical image of the WS$_2$ flakes. (b) AFM image and (c) Raman mapping of A$_{1g}$ peak mode of the corresponding areas. (d) The height profile of the red line in (b). (e) Raman spectrum of the red spot marked in (c).
}%
\end{figure}
The WS${}_2$ flakes were mechanically exfoliated from 
 bulk WS${}_2$ 
 onto silicon substrates with 300-nm silica capping layer. To verify the number of layers, Atomic force microscopy (AFM) was performed 
 with non-contact mode in air. The Raman spectrum was measured with a Raman microscope excited with a laser at $532$ nm. Microphotoluminescence ($\mu$-PL) measurements were performed at 4.2 K and pumped with a $532$-nm laser. Sample with WS$_2$ flakes was placed on 
 cryogenic confocal microscope system with high precision 3-dimensional piezo-driven motors. The PL spectra filtered with a $550$ nm long-pass filter were collected by 
 spectrometer with a 300 g/mm grating. The sample was cooled down to $4$ K with liquid helium in the cryostat with a magnetic field from $0$ to $9$ T. An electric heater was used to heat the sample from $4$ to $30$ K.

Fig. \ref{fig:geo}(a) and (b) show the optical image and AFM image of the flake with a size of a few micrometers. The Raman intensity distribution of an out-of-plane motion mode (A${}_{1g}$ peak) shown in Fig. \ref{fig:geo}(c) also reveals the shape of flakes, agreeing with optical and AFM images. According to the AFM scanning result (Fig. \ref{fig:geo}(b)), the material is not flat. The roughness of the flake surface implies the existence of defects. The result in Fig. \ref{fig:geo}(d) indicates that the height of thin part is about $1.5$ nm, which is the typical thickness of bilayer WS${}_2$\cite{WS2_atomStruture}. Although there are some contaminations in the AFM image outside the sample area (marked with dashed lines in Figure 1(b)), the bilayer height difference at the step edge is clear with considering both the AFM and Raman mode mapping. Meanwhile, according to the Raman spectrum shown in Fig. \ref{fig:geo}(e), the 2LA (second order longitude acoustic) peak at $349.13$ cm${}^{-1}$ and the A${}_{1g}$ peak at $418.58$ cm${}^{-1}$ were observed. Comparing with the Raman results with an excitation laser at $514$ nm\cite{WS2_raman}, the 2LA peak position around $349.13$ cm${}^{-1}$ is close to the peak position of trilayer and the much stronger intensity of 2LA than that of A${}_{1g}$ implies monolayer. Nevertheless, the layer number should be less than three. Combining the AFM and Raman results, it can be confirmed that the region for the optical measurement is mainly a bilayer structure.

\begin{figure}%
\centering
\includegraphics[width=\columnwidth]{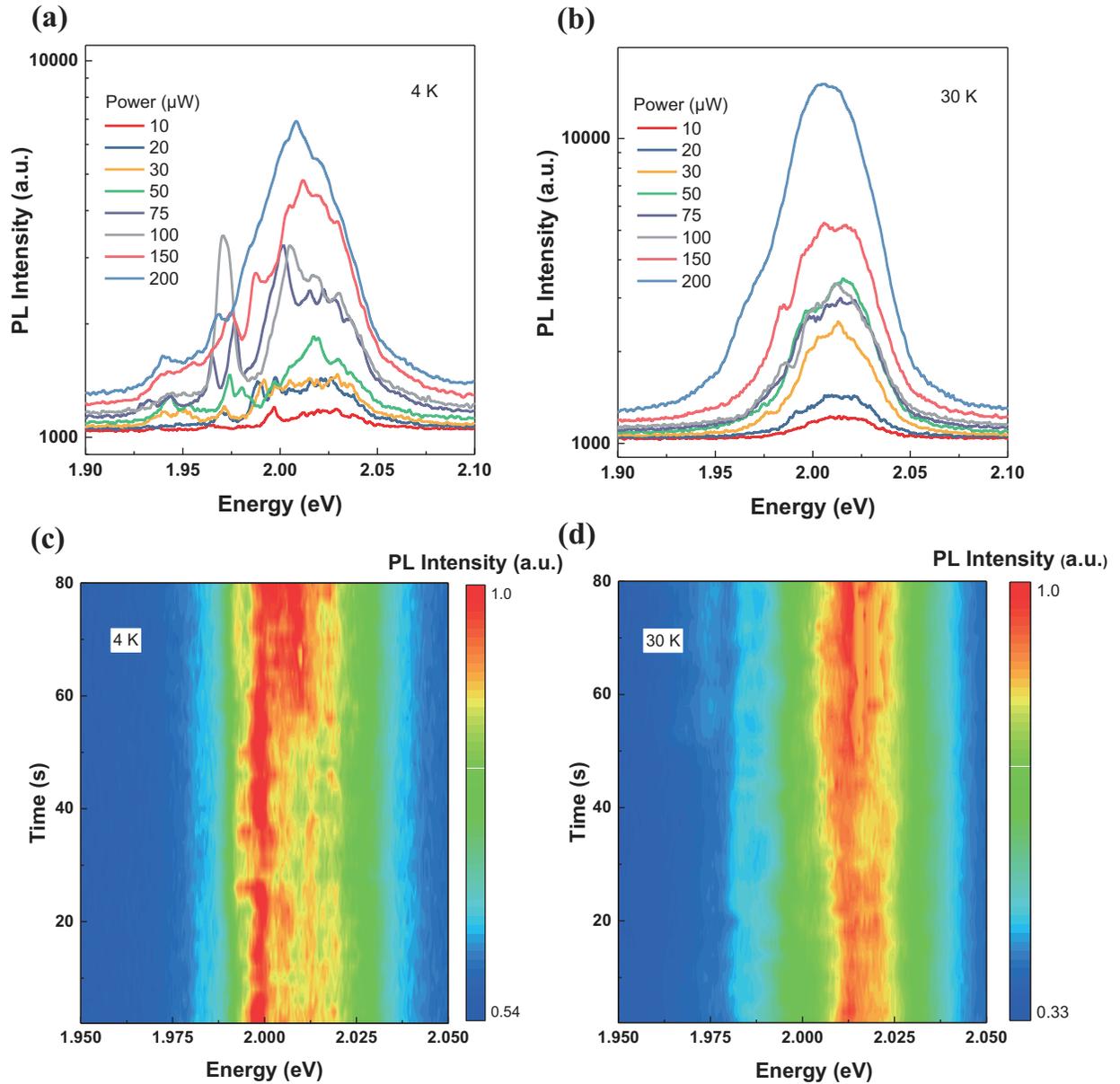}
\caption[]{\label{fig:specTime}%
\raggedright
Photoluminescence spectra at $4$ K (a) and  $30$ K (b) with increasing pumping power. (c) and (d) Contour plots of PL spectra collected continuously with an integration time of 1 s and waiting time of 1 s for a time scale of 80 s at $4$ K and $30$ K. The pumping power is kept at $100~\mu$W. The spectra at $30$ K are much more stable than those at $4$ K.
}%
\end{figure}

The measured PL spectra of our sample are 
 unstable, in which the narrow peaks emerge and disappear randomly while a relatively stable broad peak always exists, as shown in Fig. \ref{fig:specTime}. According to the peak wavelength, this wide peak is located in A exciton range, which is corresponding to the upper valence band in spin-orbit coupling\cite{xu_review}. The peak is mainly from trions rather than neutral excitons at low temperature, according to the previous temperature-dependent PL spectra\cite{temperature_1,temperature_trion}. The neutral exciton peak is much weaker than trion and defect peaks at cryogenic temperature\cite{temperature_trion}. As for the narrow peaks, not only the intensities but also the central wavelengths vary with time as shown in Fig. \ref{fig:specTime}(c) and (d). The spectral fluctuation was also observed by other groups experimentally\cite{bala_unstable,chi_defectEmission}. Since these narrow peaks located at the red side of the trion, localized defect states are the most possible sources\cite{chi_defectEmission}. In addition to the existence of fluctuation, we also discover that the fluctuation is sensitive to temperature and pumping power. The spectrum becomes more stable with increasing pumping power as shown in Fig. \ref{fig:specTime}(a) and (b). And the ratio of fluctuating peak intensity and the trion peak intensity decreases when temperature increases by comparing Fig. \ref{fig:specTime}(c) at 4.2 K  with Fig. \ref{fig:specTime}(d) at 30 K.

\begin{figure}%
\centering
\includegraphics[width=\columnwidth]{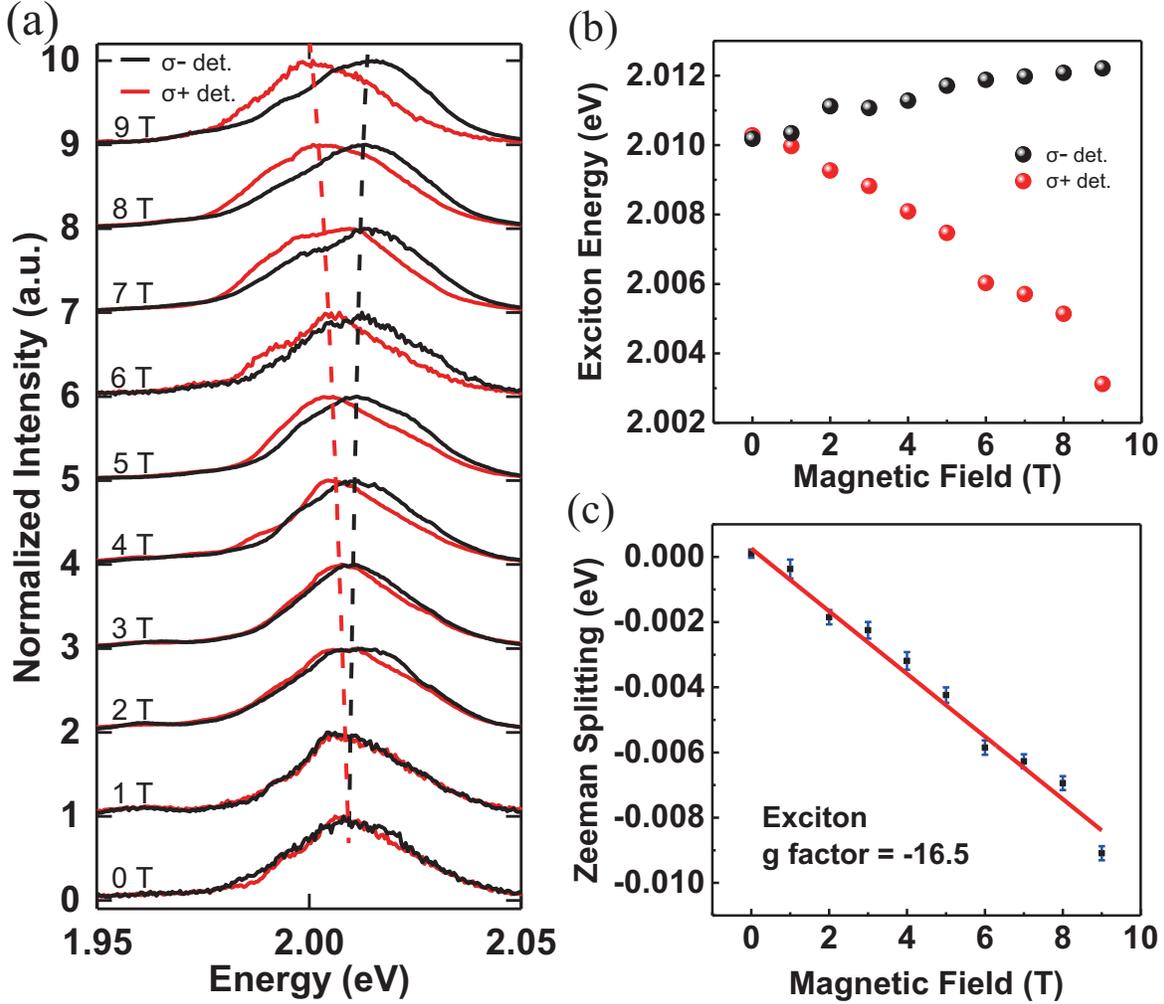}
\caption[]{\label{fig:gFactor}%
\raggedright
The Zeeman shift and the g factor at $30$ K. (a) PL spectra of two circularly polarized components with increasing magnetic field. The spectra for each Tesla are shifted for clarity. (b) and (c) The fitted peak positions and the Zeeman splittings as a function of magnetic field.
}%
\end{figure}
When a magnetic field perpendicular to the surface plane from $0$ T to $9$ T was applied, we observed a large Zeeman splitting of the wide peak as shown in Fig. \ref{fig:gFactor}(a). The sample was excited by linearly polarized laser to make two incident circularly polarized components equal. The polarization of PL was checked by a $1/4 \lambda$ wave plate and a polarizer. The $1/4 \lambda$ wave plate was controlled by a motor to measure two circularly polarized components of PL. The splitting and g factor were measured at different temperatures, ranging from $4$ K to $30$ K. Fig. \ref{fig:gFactor} (b) and (c) show the fitted peak position and Zeeman splitting respectively, with the value of Land\'{e} g factor as large as $-16.5$ at $30$ K. Additionally, there is a negative diamagnetic effect can be observed, which can be attributed to wavefunction expansion difference between initial and final states of the PL recombination for charged excitons \cite{caodiamag1,caodiamag2}.

\begin{figure}%
\centering
\includegraphics[width=\columnwidth]{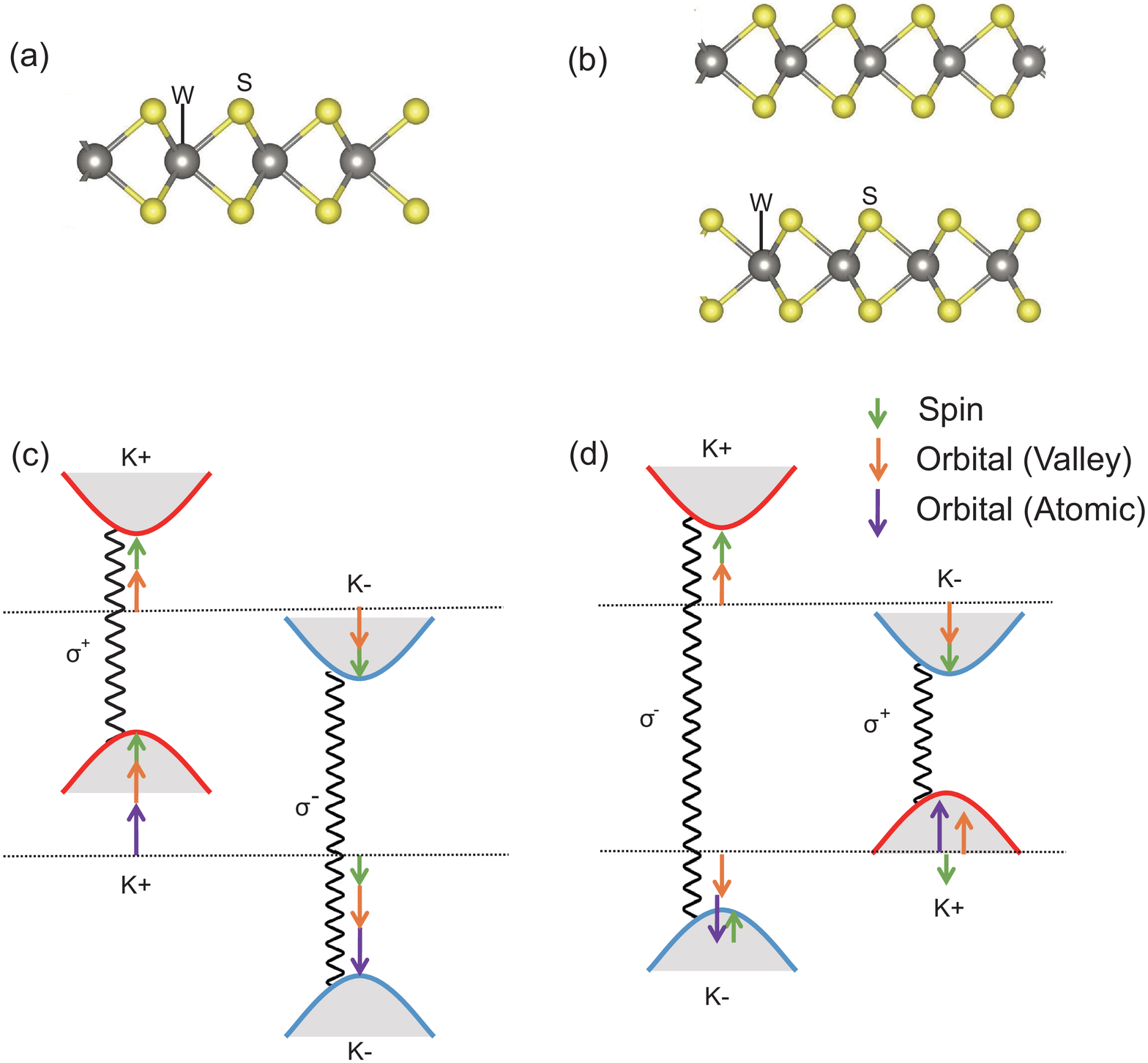}
\caption[]{\label{fig:atomBand}%
\raggedright
(a) and (b) Atom structures of monolayer and bilayer WS${}_2$. (c) and (d) Schematic energy transition diagrams of monolayer and bilayer interlayer. The arrows indicates the energy shift caused by magnetic field. The color of arrow indicates the source of magnetic moment. In monolayer situation, the shift of conduction band and valence band are in the same directions. Thus, the energy shift is relatively small. In bilayer interlayer emission, the shift of two band are in the opposite directions. Therefore, the energy shift is largely increased by magnetic field, which leads to a large g factor.
}%
\end{figure}

Figure 4 (a) and (b) show the crystal structure of single layer and bilayer WS$_2$. To explain the observed large g factor, we attribute it to the enhanced interlayer recombination of bilayer WS$_2$. The enhanced interlayer recombination could be due to the defect induced inversion symmetry breaking. Fig. \ref{fig:atomBand} depicts the energy shift contributed by each angular momentum in monolayer and interlayer hopping respectively. Early model\cite{xu_bilayerQuantumGates} of $\k \cdot \p$ theory gives the Hamiltonian as
\begin{equation}
\hat{H}_{k\cdot p} = -\epsilon_q-\lambda\tau_z\sigma_z \hat{s}_z+t_{\perp}\sigma_x + B_z \hat{s}_z.
\end{equation}
Where the Pauli matrices are defined in basis of $\Big\{\frac{1}{\sqrt{2}}\Big(|d^u_{x^2-y^2}\rangle-i\tau_z|d^u_{xy}\rangle\Big), \frac{1}{\sqrt{2}}\Big(|d^l_{x^2-y^2}\rangle+i\tau_z|d^l_{xy}\rangle\Big)\Big\}$. The superscript $u$ and $l$ mean upper and lower layer respectively. The expressions $xy,\ x^2-y^2$ describe the shapes of d orbit states of the transitional metal atom, which contribute to the states around valence band edge. $\epsilon_q$ is the energy dispersion, $\lambda$ is the spin-valley coupling, $\tau_z$ is the valley index, $t_\bot$ is the interlayer hopping for holes, and $\hat{s}_z$ denotes the spin. This model only considered spin contribution to the splitting, which is not sufficient to describe the large Zeeman splitting whose g factor is greater than $2$. The orbital contribution to the Zeeman splitting should also be considered. A precise multiband $\k \cdot \p$ model considering influences by other distant bands is more appropriate here\cite{LandauLVL_theory,multiband_theory,gFactor_6band}. The resulting g factor is given by,\cite{korn_heteroLargeG}
\begin{equation}\label{eq:gFactor}
\begin{aligned}
g &= (g_c^{\pm K} - g_v^{+K}) - (g_c^{\mp K} - g_v^{-K}),\\
  &= - 4 - 2(\frac{m_0}{m_h}\pm\frac{m_0}{m_e}).\\
\end{aligned}
\end{equation}
Where $\pm$ indicates that the transition comes from the different or same valley, $m_0$ is the mass of free electron, $m_e$ is the effective mass of electron in conduction band at $\pm$K point, and $m_h$ is the effective mass of hole in valence band at $\pm$K point. In monolayer situation, take $-$ in $\pm$, and the value of g factor is near $-4$. As for the interlayer emission, take $+$ in $\pm$, and the value of g factor is much larger. There is an equivalent explanation of the model, which is easier to understand. Zeeman splitting is contributed by three factors\cite{3comp_theory,korn_gFactorMonolayer,mac_gFactorMonolayer}, including spin, atomic orbital and valley orbital magnetic moment, as shown in Fig. \ref{fig:atomBand}(c) and (d). The spin contributes $0$ to the g factor. Because the shifts of conduction band and valence band are the same. The atomic orbital magnetic moment contributes $-4$. Because the conduction band is mainly $|d_{z^2}\rangle$, $l_z=0$. The valence band is mainly $\frac{1}{\sqrt{2}}\Big(|d_{x^2-y^2}\rangle+i\tau_z|d_{xy}\rangle\Big)$, $l_z=2\tau_z$. The valley orbital magnetic moment contributes $- 2(\frac{m_0}{m_e}\pm\frac{m_0}{m_h})$, which is related to Berry curvature\cite{andreas_magopt}. Due to spin-orbit coupling, the conduction band also slightly splits into two bands. For WS${}_2$ materials, the splitting of spin state of upper conduction band is the same with that of the upper valence band, and the lower conduction band is corresponding to the lower valence band\cite{importance_soc,3band_model_darkish_mat}.

According to the theoretical effective mass from Ref. \cite{andor_effectiveMass}, the effective electron mass $m_e$ is $0.26m_0$ for the upper conduction band and $0.35m_0$ for the lower conduction band at K point. The effective hole mass $m_h$ is $0.35m_0$ for the upper valence band and $0.49m_0$ for the lower valence band. For A exciton, only the upper valence band should be considered. For $\mathrm{WS}_2$, the spin state of upper conduction band is the same with the upper valence band\cite{importance_soc,3band_model_darkish_mat}. Therefore, for monolayer situation, the predicted value of g factor should be around $-4-2\times (1/0.35-1/0.26)=-2.0$. But for interlayer situation, the predicted value of the g factor will increase to $-4-2\times (1/0.35+1/0.26)=-17.4$, which agrees well with our experimental results. The trend that interlayer g factor is much larger than intralayer g factor is consistent with different calculation methods, although the detailed effective masses might vary \cite{andor_effectiveMass,same_mass}. If different effective masses are considered \cite{same_mass}, the calculated g factors of interlayer emission could vary in the range of $-12.9$ to $-17.4$. Nevertheless, the experimental result with value of g factor about $-16.5$ verifies our assumption that the peak mainly comes from the interlayer emission.

It should be noted that defect related bound states may also attribute to the enhancement of g factor\cite{singlePhoton_WSe2_ymh,singlePhoton_WSe2_ajit}. In our case, it cannot be totally excluded in our measurement as the PL peak position is located at energy lower than the free excitons. However, normally the defected related bound states have much narrower linewidth of PL peaks, which can be identified in our sample at low temperature and are not stable as discussed above. Therefore, we believe the interlayer emission is dominated in our measurement for the broad peak at 30K, which corresponds well with our calculations as well.

In conclusion, a value of g factor as large as $-16.5$ has been observed in a bilayer WS$_2$ at $30$ K. The large g factor is due to interlayer emission, caused by inversion symmetry breaking induced by defects in the bilayer flakes. The existence of defects has been confirmed with AFM, Raman spectroscopy and fluctuating PL peak. The large g factor has been well explained with considering the effective masses in different valleys in bilayer situation. As the g factor increases, a larger energy difference between different spin states can be obtained with a small magnetic field. The results provide a great potential for spin initialization and detection, and a platform for valleytronics for on-chip optical quantum information devices.

\begin{acknowledgments}
This work was supported by the National Natural Science Foundation of China under Grants  No. 51761145104, No. 61675228, No. 11721404 and No. 11874419; the Strategic Priority Research Program, the Instrument Developing Project and the Interdisciplinary Innovation Team of the Chinese Academy of Sciences under Grants No. XDB07030200, No. XDB28000000 and No.YJKYYQ20180036.

\end{acknowledgments}


\begin{thebibliography}{30}%
\makeatletter
\providecommand \@ifxundefined [1]{%
 \@ifx{#1\undefined}
}%
\providecommand \@ifnum [1]{%
 \ifnum #1\expandafter \@firstoftwo
 \else \expandafter \@secondoftwo
 \fi
}%
\providecommand \@ifx [1]{%
 \ifx #1\expandafter \@firstoftwo
 \else \expandafter \@secondoftwo
 \fi
}%
\providecommand \natexlab [1]{#1}%
\providecommand \enquote  [1]{``#1''}%
\providecommand \bibnamefont  [1]{#1}%
\providecommand \bibfnamefont [1]{#1}%
\providecommand \citenamefont [1]{#1}%
\providecommand \href@noop [0]{\@secondoftwo}%
\providecommand \href [0]{\begingroup \@sanitize@url \@href}%
\providecommand \@href[1]{\@@startlink{#1}\@@href}%
\providecommand \@@href[1]{\endgroup#1\@@endlink}%
\providecommand \@sanitize@url [0]{\catcode `\\12\catcode `\$12\catcode
  `\&12\catcode `\#12\catcode `\^12\catcode `\_12\catcode `\%12\relax}%
\providecommand \@@startlink[1]{}%
\providecommand \@@endlink[0]{}%
\providecommand \url  [0]{\begingroup\@sanitize@url \@url }%
\providecommand \@url [1]{\endgroup\@href {#1}{\urlprefix }}%
\providecommand \urlprefix  [0]{URL }%
\providecommand \Eprint [0]{\href }%
\providecommand \doibase [0]{http://dx.doi.org/}%
\providecommand \selectlanguage [0]{\@gobble}%
\providecommand \bibinfo  [0]{\@secondoftwo}%
\providecommand \bibfield  [0]{\@secondoftwo}%
\providecommand \translation [1]{[#1]}%
\providecommand \BibitemOpen [0]{}%
\providecommand \bibitemStop [0]{}%
\providecommand \bibitemNoStop [0]{.\EOS\space}%
\providecommand \EOS [0]{\spacefactor3000\relax}%
\providecommand \BibitemShut  [1]{\csname bibitem#1\endcsname}%
\let\auto@bib@innerbib\@empty
\bibitem [{\citenamefont {Xu}\ \emph {et~al.}(2014)\citenamefont {Xu},
  \citenamefont {Yao}, \citenamefont {Xiao},\ and\ \citenamefont
  {Heinz}}]{xu_review}%
  \BibitemOpen
  \bibfield  {author} {\bibinfo {author} {\bibfnamefont {X.~D.}\ \bibnamefont
  {Xu}}, \bibinfo {author} {\bibfnamefont {W.}~\bibnamefont {Yao}}, \bibinfo
  {author} {\bibfnamefont {D.}~\bibnamefont {Xiao}}, \ and\ \bibinfo {author}
  {\bibfnamefont {T.~F.}\ \bibnamefont {Heinz}},\ }\href@noop {} {\bibfield
  {journal} {\bibinfo  {journal} {Nat. Phys.}\ }\textbf {\bibinfo {volume}
  {10}},\ \bibinfo {pages} {343} (\bibinfo {year} {2014})}\BibitemShut
  {NoStop}%
\bibitem [{\citenamefont {Gong}\ \emph {et~al.}(2013)\citenamefont {Gong},
  \citenamefont {Liu}, \citenamefont {Yu}, \citenamefont {Xiao}, \citenamefont
  {Cui}, \citenamefont {Xu},\ and\ \citenamefont
  {Yao}}]{xu_bilayerQuantumGates}%
  \BibitemOpen
  \bibfield  {author} {\bibinfo {author} {\bibfnamefont {Z.~R.}\ \bibnamefont
  {Gong}}, \bibinfo {author} {\bibfnamefont {G.-B.}\ \bibnamefont {Liu}},
  \bibinfo {author} {\bibfnamefont {H.~Y.}\ \bibnamefont {Yu}}, \bibinfo
  {author} {\bibfnamefont {D.}~\bibnamefont {Xiao}}, \bibinfo {author}
  {\bibfnamefont {X.~D.}\ \bibnamefont {Cui}}, \bibinfo {author} {\bibfnamefont
  {X.~D.}\ \bibnamefont {Xu}}, \ and\ \bibinfo {author} {\bibfnamefont
  {W.}~\bibnamefont {Yao}},\ }\href@noop {} {\bibfield  {journal} {\bibinfo
  {journal} {Nat. Commun.}\ }\textbf {\bibinfo {volume} {4}},\ \bibinfo {pages}
  {2053} (\bibinfo {year} {2013})}\BibitemShut {NoStop}%
\bibitem [{\citenamefont {He}\ \emph {et~al.}(2015)\citenamefont {He},
  \citenamefont {Clark}, \citenamefont {Schaibley}, \citenamefont {He},
  \citenamefont {Chen}, \citenamefont {Wei}, \citenamefont {Ding},
  \citenamefont {Zhang}, \citenamefont {Yao}, \citenamefont {Xu}, \citenamefont
  {Lu},\ and\ \citenamefont {Pan}}]{singlePhoton_WSe2_ymh}%
  \BibitemOpen
  \bibfield  {author} {\bibinfo {author} {\bibfnamefont {Y.-M.}\ \bibnamefont
  {He}}, \bibinfo {author} {\bibfnamefont {G.}~\bibnamefont {Clark}}, \bibinfo
  {author} {\bibfnamefont {J.~R.}\ \bibnamefont {Schaibley}}, \bibinfo {author}
  {\bibfnamefont {Y.}~\bibnamefont {He}}, \bibinfo {author} {\bibfnamefont
  {M.-C.}\ \bibnamefont {Chen}}, \bibinfo {author} {\bibfnamefont {Y.-J.}\
  \bibnamefont {Wei}}, \bibinfo {author} {\bibfnamefont {X.}~\bibnamefont
  {Ding}}, \bibinfo {author} {\bibfnamefont {Q.}~\bibnamefont {Zhang}},
  \bibinfo {author} {\bibfnamefont {W.}~\bibnamefont {Yao}}, \bibinfo {author}
  {\bibfnamefont {X.~D.}\ \bibnamefont {Xu}}, \bibinfo {author} {\bibfnamefont
  {C.-Y.}\ \bibnamefont {Lu}}, \ and\ \bibinfo {author} {\bibfnamefont {J.-W.}\
  \bibnamefont {Pan}},\ }\href@noop {} {\bibfield  {journal} {\bibinfo
  {journal} {Nat. Nanotech.}\ }\textbf {\bibinfo {volume} {10}},\ \bibinfo
  {pages} {497} (\bibinfo {year} {2015})}\BibitemShut {NoStop}%
\bibitem [{\citenamefont {Srivastava}\ \emph {et~al.}(2015)\citenamefont
  {Srivastava}, \citenamefont {Sidler}, \citenamefont {Allain}, \citenamefont
  {Lembke}, \citenamefont {Kis},\ and\ \citenamefont
  {Imamo\v{g}lu}}]{singlePhoton_WSe2_ajit}%
  \BibitemOpen
  \bibfield  {author} {\bibinfo {author} {\bibfnamefont {A.}~\bibnamefont
  {Srivastava}}, \bibinfo {author} {\bibfnamefont {M.}~\bibnamefont {Sidler}},
  \bibinfo {author} {\bibfnamefont {A.~V.}\ \bibnamefont {Allain}}, \bibinfo
  {author} {\bibfnamefont {D.~S.}\ \bibnamefont {Lembke}}, \bibinfo {author}
  {\bibfnamefont {A.}~\bibnamefont {Kis}}, \ and\ \bibinfo {author}
  {\bibfnamefont {A.}~\bibnamefont {Imamo\v{g}lu}},\ }\href@noop {} {\bibfield
  {journal} {\bibinfo  {journal} {Nat. Nanotech.}\ }\textbf {\bibinfo {volume}
  {10}},\ \bibinfo {pages} {491} (\bibinfo {year} {2015})}\BibitemShut
  {NoStop}%
\bibitem [{\citenamefont {Koperski}\ \emph {et~al.}(2019)\citenamefont
  {Koperski}, \citenamefont {Molas}, \citenamefont {Arora}, \citenamefont
  {Nogajewski}, \citenamefont {Bartos}, \citenamefont {Wyzula}, \citenamefont
  {Vaclavkova}, \citenamefont {Kossacki},\ and\ \citenamefont
  {Potemski}}]{mac_gFactorMonolayer}%
  \BibitemOpen
  \bibfield  {author} {\bibinfo {author} {\bibfnamefont {M.}~\bibnamefont
  {Koperski}}, \bibinfo {author} {\bibfnamefont {M.~R.}\ \bibnamefont {Molas}},
  \bibinfo {author} {\bibfnamefont {A.}~\bibnamefont {Arora}}, \bibinfo
  {author} {\bibfnamefont {K.}~\bibnamefont {Nogajewski}}, \bibinfo {author}
  {\bibfnamefont {M.}~\bibnamefont {Bartos}}, \bibinfo {author} {\bibfnamefont
  {J.}~\bibnamefont {Wyzula}}, \bibinfo {author} {\bibfnamefont
  {D.}~\bibnamefont {Vaclavkova}}, \bibinfo {author} {\bibfnamefont
  {P.}~\bibnamefont {Kossacki}}, \ and\ \bibinfo {author} {\bibfnamefont
  {M.}~\bibnamefont {Potemski}},\ }\href@noop {} {\bibfield  {journal}
  {\bibinfo  {journal} {2D Materials}\ }\textbf {\bibinfo {volume} {6}},\
  \bibinfo {pages} {015001} (\bibinfo {year} {2019})}\BibitemShut {NoStop}%
\bibitem [{\citenamefont {Kuhnert}, \citenamefont {Rahimi-Iman},\ and\
  \citenamefont {Heimbrodt}(2017)}]{jan_magopt}%
  \BibitemOpen
  \bibfield  {author} {\bibinfo {author} {\bibfnamefont {J.}~\bibnamefont
  {Kuhnert}}, \bibinfo {author} {\bibfnamefont {A.}~\bibnamefont
  {Rahimi-Iman}}, \ and\ \bibinfo {author} {\bibfnamefont {W.}~\bibnamefont
  {Heimbrodt}},\ }\href@noop {} {\bibfield  {journal} {\bibinfo  {journal} {J.
  Phys.: Condens. Matter}\ }\textbf {\bibinfo {volume} {29}},\ \bibinfo {pages}
  {08LT02} (\bibinfo {year} {2017})}\BibitemShut {NoStop}%
\bibitem [{\citenamefont {Stier}\ \emph {et~al.}(2016)\citenamefont {Stier},
  \citenamefont {McCreary}, \citenamefont {Jonker}, \citenamefont {Kono},\ and\
  \citenamefont {Crooker}}]{andreas_magopt}%
  \BibitemOpen
  \bibfield  {author} {\bibinfo {author} {\bibfnamefont {A.~V.}\ \bibnamefont
  {Stier}}, \bibinfo {author} {\bibfnamefont {K.~M.}\ \bibnamefont {McCreary}},
  \bibinfo {author} {\bibfnamefont {B.~T.}\ \bibnamefont {Jonker}}, \bibinfo
  {author} {\bibfnamefont {J.}~\bibnamefont {Kono}}, \ and\ \bibinfo {author}
  {\bibfnamefont {S.~A.}\ \bibnamefont {Crooker}},\ }\href@noop {} {\bibfield
  {journal} {\bibinfo  {journal} {Nat. Commun.}\ }\textbf {\bibinfo {volume}
  {7}},\ \bibinfo {pages} {10643} (\bibinfo {year} {2016})}\BibitemShut
  {NoStop}%
\bibitem [{\citenamefont {Plechinger}\ \emph {et~al.}(2016)\citenamefont
  {Plechinger}, \citenamefont {Nagler}, \citenamefont {Arora}, \citenamefont
  {Aguila}, \citenamefont {Ballottin}, \citenamefont {Frank}, \citenamefont
  {Steinleitner}, \citenamefont {Gmitra}, \citenamefont {Fabian}, \citenamefont
  {Christianen}, \citenamefont {Bratschitsch}, \citenamefont {Sch\"{u}ller},\
  and\ \citenamefont {Korn}}]{korn_gFactorMonolayer}%
  \BibitemOpen
  \bibfield  {author} {\bibinfo {author} {\bibfnamefont {G.}~\bibnamefont
  {Plechinger}}, \bibinfo {author} {\bibfnamefont {P.}~\bibnamefont {Nagler}},
  \bibinfo {author} {\bibfnamefont {A.}~\bibnamefont {Arora}}, \bibinfo
  {author} {\bibfnamefont {A.}~\bibnamefont {Aguila}}, \bibinfo {author}
  {\bibfnamefont {M.}~\bibnamefont {Ballottin}}, \bibinfo {author}
  {\bibfnamefont {T.}~\bibnamefont {Frank}}, \bibinfo {author} {\bibfnamefont
  {P.}~\bibnamefont {Steinleitner}}, \bibinfo {author} {\bibfnamefont
  {M.}~\bibnamefont {Gmitra}}, \bibinfo {author} {\bibfnamefont
  {J.}~\bibnamefont {Fabian}}, \bibinfo {author} {\bibfnamefont
  {P.}~\bibnamefont {Christianen}}, \bibinfo {author} {\bibfnamefont
  {R.}~\bibnamefont {Bratschitsch}}, \bibinfo {author} {\bibfnamefont
  {C.}~\bibnamefont {Sch\"{u}ller}}, \ and\ \bibinfo {author} {\bibfnamefont
  {T.}~\bibnamefont {Korn}},\ }\href@noop {} {\bibfield  {journal} {\bibinfo
  {journal} {Nano Lett.}\ }\textbf {\bibinfo {volume} {16}},\ \bibinfo {pages}
  {7899} (\bibinfo {year} {2016})}\BibitemShut {NoStop}%
\bibitem [{\citenamefont {Wang}, \citenamefont {Mak},\ and\ \citenamefont
  {Shan}(2018)}]{PhysRevLett.120.066402}%
  \BibitemOpen
  \bibfield  {author} {\bibinfo {author} {\bibfnamefont {Z.}~\bibnamefont
  {Wang}}, \bibinfo {author} {\bibfnamefont {K.~F.}\ \bibnamefont {Mak}}, \
  and\ \bibinfo {author} {\bibfnamefont {J.}~\bibnamefont {Shan}},\ }\href
  {\doibase 10.1103/PhysRevLett.120.066402} {\bibfield  {journal} {\bibinfo
  {journal} {Phys. Rev. Lett.}\ }\textbf {\bibinfo {volume} {120}},\ \bibinfo
  {pages} {066402} (\bibinfo {year} {2018})}\BibitemShut {NoStop}%
\bibitem [{\citenamefont {Back}\ \emph {et~al.}(2017)\citenamefont {Back},
  \citenamefont {Sidler}, \citenamefont {Cotlet}, \citenamefont {Srivastava},
  \citenamefont {Takemura}, \citenamefont {Kroner},\ and\ \citenamefont
  {Imamo\v{g}lu}}]{gFactor38_MoSe2}%
  \BibitemOpen
  \bibfield  {author} {\bibinfo {author} {\bibfnamefont {P.}~\bibnamefont
  {Back}}, \bibinfo {author} {\bibfnamefont {M.}~\bibnamefont {Sidler}},
  \bibinfo {author} {\bibfnamefont {O.}~\bibnamefont {Cotlet}}, \bibinfo
  {author} {\bibfnamefont {A.}~\bibnamefont {Srivastava}}, \bibinfo {author}
  {\bibfnamefont {N.}~\bibnamefont {Takemura}}, \bibinfo {author}
  {\bibfnamefont {M.}~\bibnamefont {Kroner}}, \ and\ \bibinfo {author}
  {\bibfnamefont {A.}~\bibnamefont {Imamo\v{g}lu}},\ }\href@noop {} {\bibfield
  {journal} {\bibinfo  {journal} {Phys. Rev. Lett.}\ }\textbf {\bibinfo
  {volume} {118}},\ \bibinfo {pages} {237404} (\bibinfo {year}
  {2017})}\BibitemShut {NoStop}%
\bibitem [{\citenamefont {Gustafsson}\ \emph {et~al.}(2018)\citenamefont
  {Gustafsson}, \citenamefont {Yankowitz}, \citenamefont {Forsythe},
  \citenamefont {Rhodes}, \citenamefont {Watanabe}, \citenamefont {Taniguchi},
  \citenamefont {Hone}, \citenamefont {Zhu},\ and\ \citenamefont
  {Dean}}]{Gustafsson2018}%
  \BibitemOpen
  \bibfield  {author} {\bibinfo {author} {\bibfnamefont {M.~V.}\ \bibnamefont
  {Gustafsson}}, \bibinfo {author} {\bibfnamefont {M.}~\bibnamefont
  {Yankowitz}}, \bibinfo {author} {\bibfnamefont {C.}~\bibnamefont {Forsythe}},
  \bibinfo {author} {\bibfnamefont {D.}~\bibnamefont {Rhodes}}, \bibinfo
  {author} {\bibfnamefont {K.}~\bibnamefont {Watanabe}}, \bibinfo {author}
  {\bibfnamefont {T.}~\bibnamefont {Taniguchi}}, \bibinfo {author}
  {\bibfnamefont {J.}~\bibnamefont {Hone}}, \bibinfo {author} {\bibfnamefont
  {X.}~\bibnamefont {Zhu}}, \ and\ \bibinfo {author} {\bibfnamefont {C.~R.}\
  \bibnamefont {Dean}},\ }\href {\doibase 10.1038/s41563-018-0036-2} {\bibfield
   {journal} {\bibinfo  {journal} {Nature Materials}\ }\textbf {\bibinfo
  {volume} {17}},\ \bibinfo {pages} {411} (\bibinfo {year} {2018})}\BibitemShut
  {NoStop}%
\bibitem [{\citenamefont {Zhu}\ \emph {et~al.}(2014)\citenamefont {Zhu},
  \citenamefont {Zeng}, \citenamefont {Dai}, \citenamefont {Gong},\ and\
  \citenamefont {Cui}}]{cui_forbidInterlayer}%
  \BibitemOpen
  \bibfield  {author} {\bibinfo {author} {\bibfnamefont {B.~R.}\ \bibnamefont
  {Zhu}}, \bibinfo {author} {\bibfnamefont {H.~L.}\ \bibnamefont {Zeng}},
  \bibinfo {author} {\bibfnamefont {J.~F.}\ \bibnamefont {Dai}}, \bibinfo
  {author} {\bibfnamefont {Z.~R.}\ \bibnamefont {Gong}}, \ and\ \bibinfo
  {author} {\bibfnamefont {X.~D.}\ \bibnamefont {Cui}},\ }\href@noop {}
  {\bibfield  {journal} {\bibinfo  {journal} {PNAS}\ }\textbf {\bibinfo
  {volume} {111}},\ \bibinfo {pages} {11606} (\bibinfo {year}
  {2014})}\BibitemShut {NoStop}%
\bibitem [{\citenamefont {Nagler}\ \emph {et~al.}(2017)\citenamefont {Nagler},
  \citenamefont {Ballottin}, \citenamefont {Mitioglu}, \citenamefont
  {Mooshammer}, \citenamefont {Paradiso}, \citenamefont {Strunk}, \citenamefont
  {Huber}, \citenamefont {Chernikov}, \citenamefont {Christianen},
  \citenamefont {Sch\"{u}ller},\ and\ \citenamefont
  {Korn}}]{korn_heteroLargeG}%
  \BibitemOpen
  \bibfield  {author} {\bibinfo {author} {\bibfnamefont {P.}~\bibnamefont
  {Nagler}}, \bibinfo {author} {\bibfnamefont {M.~V.}\ \bibnamefont
  {Ballottin}}, \bibinfo {author} {\bibfnamefont {A.~A.}\ \bibnamefont
  {Mitioglu}}, \bibinfo {author} {\bibfnamefont {F.}~\bibnamefont
  {Mooshammer}}, \bibinfo {author} {\bibfnamefont {N.}~\bibnamefont
  {Paradiso}}, \bibinfo {author} {\bibfnamefont {C.}~\bibnamefont {Strunk}},
  \bibinfo {author} {\bibfnamefont {R.}~\bibnamefont {Huber}}, \bibinfo
  {author} {\bibfnamefont {A.}~\bibnamefont {Chernikov}}, \bibinfo {author}
  {\bibfnamefont {P.~C.~M.}\ \bibnamefont {Christianen}}, \bibinfo {author}
  {\bibfnamefont {C.}~\bibnamefont {Sch\"{u}ller}}, \ and\ \bibinfo {author}
  {\bibfnamefont {T.}~\bibnamefont {Korn}},\ }\href@noop {} {\bibfield
  {journal} {\bibinfo  {journal} {Nat. Commun.}\ }\textbf {\bibinfo {volume}
  {8}},\ \bibinfo {pages} {1551} (\bibinfo {year} {2017})}\BibitemShut
  {NoStop}%
\bibitem [{\citenamefont {Jiang}\ \emph {et~al.}(2017)\citenamefont {Jiang},
  \citenamefont {Liu}, \citenamefont {Cuadra}, \citenamefont {Huang},
  \citenamefont {Li}, \citenamefont {Srivastava}, \citenamefont {Liu},\ and\
  \citenamefont {Gao}}]{gao_gFactor}%
  \BibitemOpen
  \bibfield  {author} {\bibinfo {author} {\bibfnamefont {C.~Y.}\ \bibnamefont
  {Jiang}}, \bibinfo {author} {\bibfnamefont {F.~C.}\ \bibnamefont {Liu}},
  \bibinfo {author} {\bibfnamefont {J.}~\bibnamefont {Cuadra}}, \bibinfo
  {author} {\bibfnamefont {Z.~M.}\ \bibnamefont {Huang}}, \bibinfo {author}
  {\bibfnamefont {K.}~\bibnamefont {Li}}, \bibinfo {author} {\bibfnamefont
  {A.}~\bibnamefont {Srivastava}}, \bibinfo {author} {\bibfnamefont
  {Z.}~\bibnamefont {Liu}}, \ and\ \bibinfo {author} {\bibfnamefont {W.-B.}\
  \bibnamefont {Gao}},\ }\href@noop {} {\bibfield  {journal} {\bibinfo
  {journal} {Nat. Commun.}\ }\textbf {\bibinfo {volume} {8}},\ \bibinfo {pages}
  {802} (\bibinfo {year} {2017})}\BibitemShut {NoStop}%
\bibitem [{\citenamefont {Persson}(2016)}]{WS2_atomStruture}%
  \BibitemOpen
  \bibfield  {author} {\bibinfo {author} {\bibfnamefont {K.}~\bibnamefont
  {Persson}},\ }\href@noop {} {\enquote {\bibinfo {title} {Materials data on
  ws2 (sg:194)},}\ }\bibinfo {howpublished} {https://materialsproject.org}
  (\bibinfo {year} {2016})\BibitemShut {NoStop}%
\bibitem [{\citenamefont {Berkdemir}\ \emph {et~al.}(2013)\citenamefont
  {Berkdemir}, \citenamefont {Guti\'{e}rrez}, \citenamefont
  {Botello-M\'{e}ndez}, \citenamefont {P.-L\'{o}pez}, \citenamefont
  {El\'{i}as}, \citenamefont {Chia}, \citenamefont {Wang}, \citenamefont
  {Crespi}, \citenamefont {L.-Ur\'{i}as}, \citenamefont {Charlier},
  \citenamefont {Terrones},\ and\ \citenamefont {Terrones}}]{WS2_raman}%
  \BibitemOpen
  \bibfield  {author} {\bibinfo {author} {\bibfnamefont {A.}~\bibnamefont
  {Berkdemir}}, \bibinfo {author} {\bibfnamefont {H.~R.}\ \bibnamefont
  {Guti\'{e}rrez}}, \bibinfo {author} {\bibfnamefont {A.~R.}\ \bibnamefont
  {Botello-M\'{e}ndez}}, \bibinfo {author} {\bibfnamefont {N.}~\bibnamefont
  {P.-L\'{o}pez}}, \bibinfo {author} {\bibfnamefont {A.~L.}\ \bibnamefont
  {El\'{i}as}}, \bibinfo {author} {\bibfnamefont {C.-I.}\ \bibnamefont {Chia}},
  \bibinfo {author} {\bibfnamefont {B.}~\bibnamefont {Wang}}, \bibinfo {author}
  {\bibfnamefont {V.~H.}\ \bibnamefont {Crespi}}, \bibinfo {author}
  {\bibfnamefont {F.}~\bibnamefont {L.-Ur\'{i}as}}, \bibinfo {author}
  {\bibfnamefont {J.-C.}\ \bibnamefont {Charlier}}, \bibinfo {author}
  {\bibfnamefont {H.}~\bibnamefont {Terrones}}, \ and\ \bibinfo {author}
  {\bibfnamefont {M.}~\bibnamefont {Terrones}},\ }\href@noop {} {\bibfield
  {journal} {\bibinfo  {journal} {Sci. Rep.}\ }\textbf {\bibinfo {volume}
  {3}},\ \bibinfo {pages} {1755} (\bibinfo {year} {2013})}\BibitemShut
  {NoStop}%
\bibitem [{\citenamefont {Kato}\ and\ \citenamefont
  {Kaneko}(2016)}]{temperature_1}%
  \BibitemOpen
  \bibfield  {author} {\bibinfo {author} {\bibfnamefont {T.}~\bibnamefont
  {Kato}}\ and\ \bibinfo {author} {\bibfnamefont {T.}~\bibnamefont {Kaneko}},\
  }\href@noop {} {\bibfield  {journal} {\bibinfo  {journal} {ACS Nano}\
  }\textbf {\bibinfo {volume} {10}},\ \bibinfo {pages} {9687} (\bibinfo {year}
  {2016})}\BibitemShut {NoStop}%
\bibitem [{\citenamefont {Plechinger}\ \emph {et~al.}(2015)\citenamefont
  {Plechinger}, \citenamefont {Nagler}, \citenamefont {Kraus}, \citenamefont
  {Paradiso}, \citenamefont {Strunk}, \citenamefont {Sch\"{u}ller},\ and\
  \citenamefont {Korn}}]{temperature_trion}%
  \BibitemOpen
  \bibfield  {author} {\bibinfo {author} {\bibfnamefont {G.}~\bibnamefont
  {Plechinger}}, \bibinfo {author} {\bibfnamefont {P.}~\bibnamefont {Nagler}},
  \bibinfo {author} {\bibfnamefont {J.}~\bibnamefont {Kraus}}, \bibinfo
  {author} {\bibfnamefont {N.}~\bibnamefont {Paradiso}}, \bibinfo {author}
  {\bibfnamefont {C.}~\bibnamefont {Strunk}}, \bibinfo {author} {\bibfnamefont
  {C.}~\bibnamefont {Sch\"{u}ller}}, \ and\ \bibinfo {author} {\bibfnamefont
  {T.}~\bibnamefont {Korn}},\ }\href@noop {} {\bibfield  {journal} {\bibinfo
  {journal} {Phys. Stat. Sol. (RRL)}\ }\textbf {\bibinfo {volume} {9}},\
  \bibinfo {pages} {457} (\bibinfo {year} {2015})}\BibitemShut {NoStop}%
\bibitem [{\citenamefont {Bala}\ \emph {et~al.}(2016)\citenamefont {Bala},
  \citenamefont {\L{}aci\'{n}ska}, \citenamefont {Nogajewski}, \citenamefont
  {Molas}, \citenamefont {Wysmo\l{}ek},\ and\ \citenamefont
  {Potemski}}]{bala_unstable}%
  \BibitemOpen
  \bibfield  {author} {\bibinfo {author} {\bibfnamefont {L.}~\bibnamefont
  {Bala}}, \bibinfo {author} {\bibfnamefont {E.~M.}\ \bibnamefont
  {\L{}aci\'{n}ska}}, \bibinfo {author} {\bibfnamefont {K.}~\bibnamefont
  {Nogajewski}}, \bibinfo {author} {\bibfnamefont {M.~R.}\ \bibnamefont
  {Molas}}, \bibinfo {author} {\bibfnamefont {A.}~\bibnamefont {Wysmo\l{}ek}},
  \ and\ \bibinfo {author} {\bibfnamefont {M.}~\bibnamefont {Potemski}},\
  }\href@noop {} {\bibfield  {journal} {\bibinfo  {journal} {Acta Phys. Pol.
  A}\ }\textbf {\bibinfo {volume} {5}},\ \bibinfo {pages} {1176} (\bibinfo
  {year} {2016})}\BibitemShut {NoStop}%
\bibitem [{\citenamefont {Chakraborty}, \citenamefont {Goodfellow},\ and\
  \citenamefont {Vamivakas}(2016)}]{chi_defectEmission}%
  \BibitemOpen
  \bibfield  {author} {\bibinfo {author} {\bibfnamefont {C.}~\bibnamefont
  {Chakraborty}}, \bibinfo {author} {\bibfnamefont {K.~M.}\ \bibnamefont
  {Goodfellow}}, \ and\ \bibinfo {author} {\bibfnamefont {A.~N.}\ \bibnamefont
  {Vamivakas}},\ }\href@noop {} {\bibfield  {journal} {\bibinfo  {journal}
  {Opt. Mater. Express}\ }\textbf {\bibinfo {volume} {6}},\ \bibinfo {pages}
  {2081} (\bibinfo {year} {2016})}\BibitemShut {NoStop}%
\bibitem [{\citenamefont {Cao}\ \emph {et~al.}(2015)\citenamefont {Cao},
  \citenamefont {Tang}, \citenamefont {Gao}, \citenamefont {Sun}, \citenamefont
  {Qiu}, \citenamefont {Zhao}, \citenamefont {He}, \citenamefont {Shi},
  \citenamefont {Gu}, \citenamefont {Williams} \emph {et~al.}}]{caodiamag1}%
  \BibitemOpen
  \bibfield  {author} {\bibinfo {author} {\bibfnamefont {S.}~\bibnamefont
  {Cao}}, \bibinfo {author} {\bibfnamefont {J.}~\bibnamefont {Tang}}, \bibinfo
  {author} {\bibfnamefont {Y.}~\bibnamefont {Gao}}, \bibinfo {author}
  {\bibfnamefont {Y.}~\bibnamefont {Sun}}, \bibinfo {author} {\bibfnamefont
  {K.}~\bibnamefont {Qiu}}, \bibinfo {author} {\bibfnamefont {Y.}~\bibnamefont
  {Zhao}}, \bibinfo {author} {\bibfnamefont {M.}~\bibnamefont {He}}, \bibinfo
  {author} {\bibfnamefont {J.-A.}\ \bibnamefont {Shi}}, \bibinfo {author}
  {\bibfnamefont {L.}~\bibnamefont {Gu}}, \bibinfo {author} {\bibfnamefont
  {D.~A.}\ \bibnamefont {Williams}},  \emph {et~al.},\ }\href@noop {}
  {\bibfield  {journal} {\bibinfo  {journal} {Scientific reports}\ }\textbf
  {\bibinfo {volume} {5}},\ \bibinfo {pages} {8041} (\bibinfo {year}
  {2015})}\BibitemShut {NoStop}%
\bibitem [{\citenamefont {Cao}\ \emph {et~al.}(2016)\citenamefont {Cao},
  \citenamefont {Tang}, \citenamefont {Sun}, \citenamefont {Peng},
  \citenamefont {Gao}, \citenamefont {Zhao}, \citenamefont {Qian},
  \citenamefont {Sun}, \citenamefont {Ali}, \citenamefont {Shao} \emph
  {et~al.}}]{caodiamag2}%
  \BibitemOpen
  \bibfield  {author} {\bibinfo {author} {\bibfnamefont {S.}~\bibnamefont
  {Cao}}, \bibinfo {author} {\bibfnamefont {J.}~\bibnamefont {Tang}}, \bibinfo
  {author} {\bibfnamefont {Y.}~\bibnamefont {Sun}}, \bibinfo {author}
  {\bibfnamefont {K.}~\bibnamefont {Peng}}, \bibinfo {author} {\bibfnamefont
  {Y.}~\bibnamefont {Gao}}, \bibinfo {author} {\bibfnamefont {Y.}~\bibnamefont
  {Zhao}}, \bibinfo {author} {\bibfnamefont {C.}~\bibnamefont {Qian}}, \bibinfo
  {author} {\bibfnamefont {S.}~\bibnamefont {Sun}}, \bibinfo {author}
  {\bibfnamefont {H.}~\bibnamefont {Ali}}, \bibinfo {author} {\bibfnamefont
  {Y.}~\bibnamefont {Shao}},  \emph {et~al.},\ }\href@noop {} {\bibfield
  {journal} {\bibinfo  {journal} {Nano Research}\ }\textbf {\bibinfo {volume}
  {9}},\ \bibinfo {pages} {306} (\bibinfo {year} {2016})}\BibitemShut {NoStop}%
\bibitem [{\citenamefont {Korm{\'a}nyos}, \citenamefont {Rakyta},\ and\
  \citenamefont {Burkard}(2015)}]{LandauLVL_theory}%
  \BibitemOpen
  \bibfield  {author} {\bibinfo {author} {\bibfnamefont {A.}~\bibnamefont
  {Korm{\'a}nyos}}, \bibinfo {author} {\bibfnamefont {P.}~\bibnamefont
  {Rakyta}}, \ and\ \bibinfo {author} {\bibfnamefont {G.}~\bibnamefont
  {Burkard}},\ }\href@noop {} {\bibfield  {journal} {\bibinfo  {journal} {New
  Journal of Physics}\ }\textbf {\bibinfo {volume} {17}},\ \bibinfo {pages}
  {103006} (\bibinfo {year} {2015})}\BibitemShut {NoStop}%
\bibitem [{\citenamefont {Wang}\ \emph {et~al.}(2015)\citenamefont {Wang},
  \citenamefont {Bouet}, \citenamefont {Glazov}, \citenamefont {Ivchenko},
  \citenamefont {Palleau}, \citenamefont {Marie},\ and\ \citenamefont
  {Urbaszek}}]{multiband_theory}%
  \BibitemOpen
  \bibfield  {author} {\bibinfo {author} {\bibfnamefont {G.}~\bibnamefont
  {Wang}}, \bibinfo {author} {\bibfnamefont {L.}~\bibnamefont {Bouet}},
  \bibinfo {author} {\bibfnamefont {M.~M.}\ \bibnamefont {Glazov}}, \bibinfo
  {author} {\bibfnamefont {T.~A. E.~L.}\ \bibnamefont {Ivchenko}}, \bibinfo
  {author} {\bibfnamefont {E.}~\bibnamefont {Palleau}}, \bibinfo {author}
  {\bibfnamefont {X.}~\bibnamefont {Marie}}, \ and\ \bibinfo {author}
  {\bibfnamefont {B.}~\bibnamefont {Urbaszek}},\ }\href@noop {} {\bibfield
  {journal} {\bibinfo  {journal} {2D Materials}\ }\textbf {\bibinfo {volume}
  {2}},\ \bibinfo {pages} {034002} (\bibinfo {year} {2015})}\BibitemShut
  {NoStop}%
\bibitem [{\citenamefont {Rybkovskiy}, \citenamefont {Gerber},\ and\
  \citenamefont {Durnev}(2017)}]{gFactor_6band}%
  \BibitemOpen
  \bibfield  {author} {\bibinfo {author} {\bibfnamefont {D.~V.}\ \bibnamefont
  {Rybkovskiy}}, \bibinfo {author} {\bibfnamefont {I.~C.}\ \bibnamefont
  {Gerber}}, \ and\ \bibinfo {author} {\bibfnamefont {M.~V.}\ \bibnamefont
  {Durnev}},\ }\href@noop {} {\bibfield  {journal} {\bibinfo  {journal} {Phys.
  Rev. B}\ }\textbf {\bibinfo {volume} {95}},\ \bibinfo {pages} {155406}
  (\bibinfo {year} {2017})}\BibitemShut {NoStop}%
\bibitem [{\citenamefont {Koperski}\ \emph {et~al.}(2017)\citenamefont
  {Koperski}, \citenamefont {Molas}, \citenamefont {Arora}, \citenamefont
  {Nogajewski}, \citenamefont {Slobodeniuk}, \citenamefont {Faugeras},\ and\
  \citenamefont {Potemski}}]{3comp_theory}%
  \BibitemOpen
  \bibfield  {author} {\bibinfo {author} {\bibfnamefont {M.}~\bibnamefont
  {Koperski}}, \bibinfo {author} {\bibfnamefont {M.~R.}\ \bibnamefont {Molas}},
  \bibinfo {author} {\bibfnamefont {A.}~\bibnamefont {Arora}}, \bibinfo
  {author} {\bibfnamefont {K.}~\bibnamefont {Nogajewski}}, \bibinfo {author}
  {\bibfnamefont {A.~O.}\ \bibnamefont {Slobodeniuk}}, \bibinfo {author}
  {\bibfnamefont {C.}~\bibnamefont {Faugeras}}, \ and\ \bibinfo {author}
  {\bibfnamefont {M.}~\bibnamefont {Potemski}},\ }\href@noop {} {\bibfield
  {journal} {\bibinfo  {journal} {Nanophotonics}\ }\textbf {\bibinfo {volume}
  {6}},\ \bibinfo {pages} {1289} (\bibinfo {year} {2017})}\BibitemShut
  {NoStop}%
\bibitem [{\citenamefont {Korm\'{a}nyos}\ \emph {et~al.}(2013)\citenamefont
  {Korm\'{a}nyos}, \citenamefont {Z\'{o}lyomi}, \citenamefont {Drummond},
  \citenamefont {Rakyta}, \citenamefont {Burkard},\ and\ \citenamefont
  {Fal'ko}}]{importance_soc}%
  \BibitemOpen
  \bibfield  {author} {\bibinfo {author} {\bibfnamefont {A.}~\bibnamefont
  {Korm\'{a}nyos}}, \bibinfo {author} {\bibfnamefont {V.}~\bibnamefont
  {Z\'{o}lyomi}}, \bibinfo {author} {\bibfnamefont {N.~D.}\ \bibnamefont
  {Drummond}}, \bibinfo {author} {\bibfnamefont {P.}~\bibnamefont {Rakyta}},
  \bibinfo {author} {\bibfnamefont {G.}~\bibnamefont {Burkard}}, \ and\
  \bibinfo {author} {\bibfnamefont {V.~I.}\ \bibnamefont {Fal'ko}},\
  }\href@noop {} {\bibfield  {journal} {\bibinfo  {journal} {Phys. Rev. B}\
  }\textbf {\bibinfo {volume} {88}},\ \bibinfo {pages} {045416} (\bibinfo
  {year} {2013})}\BibitemShut {NoStop}%
\bibitem [{\citenamefont {Liu}\ \emph {et~al.}(2013)\citenamefont {Liu},
  \citenamefont {Shan}, \citenamefont {Yao}, \citenamefont {Yao},\ and\
  \citenamefont {Xiao}}]{3band_model_darkish_mat}%
  \BibitemOpen
  \bibfield  {author} {\bibinfo {author} {\bibfnamefont {G.-B.}\ \bibnamefont
  {Liu}}, \bibinfo {author} {\bibfnamefont {W.-Y.}\ \bibnamefont {Shan}},
  \bibinfo {author} {\bibfnamefont {Y.~G.}\ \bibnamefont {Yao}}, \bibinfo
  {author} {\bibfnamefont {W.}~\bibnamefont {Yao}}, \ and\ \bibinfo {author}
  {\bibfnamefont {D.}~\bibnamefont {Xiao}},\ }\href@noop {} {\bibfield
  {journal} {\bibinfo  {journal} {Phys. Rev. B}\ }\textbf {\bibinfo {volume}
  {88}},\ \bibinfo {pages} {085433} (\bibinfo {year} {2013})}\BibitemShut
  {NoStop}%
\bibitem [{\citenamefont {Korm\'{a}nyos}\ \emph {et~al.}(2015)\citenamefont
  {Korm\'{a}nyos}, \citenamefont {Burkard}, \citenamefont {Gmitra},
  \citenamefont {Fabian}, \citenamefont {Z\'{o}lyomi}, \citenamefont
  {Drummond},\ and\ \citenamefont {Fal'ko}}]{andor_effectiveMass}%
  \BibitemOpen
  \bibfield  {author} {\bibinfo {author} {\bibfnamefont {A.}~\bibnamefont
  {Korm\'{a}nyos}}, \bibinfo {author} {\bibfnamefont {G.}~\bibnamefont
  {Burkard}}, \bibinfo {author} {\bibfnamefont {M.}~\bibnamefont {Gmitra}},
  \bibinfo {author} {\bibfnamefont {J.}~\bibnamefont {Fabian}}, \bibinfo
  {author} {\bibfnamefont {V.}~\bibnamefont {Z\'{o}lyomi}}, \bibinfo {author}
  {\bibfnamefont {N.~D.}\ \bibnamefont {Drummond}}, \ and\ \bibinfo {author}
  {\bibfnamefont {V.}~\bibnamefont {Fal'ko}},\ }\href@noop {} {\bibfield
  {journal} {\bibinfo  {journal} {2D Materials}\ }\textbf {\bibinfo {volume}
  {2}},\ \bibinfo {pages} {022001} (\bibinfo {year} {2015})}\BibitemShut
  {NoStop}%
\bibitem [{\citenamefont {Ramasubramaniam}(2012)}]{same_mass}%
  \BibitemOpen
  \bibfield  {author} {\bibinfo {author} {\bibfnamefont {A.}~\bibnamefont
  {Ramasubramaniam}},\ }\href@noop {} {\bibfield  {journal} {\bibinfo
  {journal} {Phys. Rev. B}\ }\textbf {\bibinfo {volume} {86}},\ \bibinfo
  {pages} {115409} (\bibinfo {year} {2012})}\BibitemShut {NoStop}%
\end{thebibliography}
\end{document}